\documentclass[aps,twocolumn,groupedaddress,showpacs,floats]{revtex4}
\usepackage[dvips]{graphicx}
\usepackage{bm}
\begin{document}

\title{Commensurate Two-Component Bosons in Optical Lattice:\\ Groundstate Phase Diagram}
\author{Anatoly Kuklov${^1}$,
Nikolay Prokof'ev$^{2,3}$, and Boris Svistunov$^{2,3}$}
\affiliation{ ${^1}$ Department of Engineering Science and
Physics, The College of Staten Island, City University of New
York, Staten
Island, New York 10314  \\
${^2}$ Department of Physics, University of
Massachusetts, Amherst, MA 01003 \\
${^3}$ Russian Research Center ``Kurchatov Institute'',
123182 Moscow
}

\begin{abstract}
Two sorts of bosons in an optical lattice at commensurate
filling factors can form five stable superfluid and insulating
groundstates with rich and non-trivial phase diagram.
The structure of the groundstate diagram is established
by mapping $d$-dimensional quantum system onto a $(d+1)$-dimensional
classical loop-current model and Monte Carlo (MC) simulations of the
latter. Surprisingly, the quantum phase diagram features,
besides second-order lines, first-order transitions and two
multi-critical points.
We explain why first-order transitions are generic for models
with pairing interactions using microscopic and mean-field (MF) arguments.
In some cases, the MC results strongly deviate from the MF predictions.

\end{abstract}

\pacs{03.75.Kk, 05.30.Jp}

\maketitle

Ultracold atoms trapped in an optical lattice (OL)
\cite{Jaksch,Greiner} form an intriguing strongly correlated
quantum system. The unprecedented control over parameters of the
effective Hubbard-type Hamiltonian renders this system an
important object for the study of quantum phase transitions
\cite{Sachdev}. Single-component bosons without internal degrees
of freedom have only two phases in a regular lattice: superfluid
(SF) and Mott-insulator (MI) (at 
commensurate filling factor \cite{Fisher}). When several bosonic
species are combined in the OL, the  na\"ive expectation that
their groundstates are straightforward mixtures of MI and SF with
respect to participating components is wrong---the phases of
spinor and multi-component systems are far more subtle
\cite{Demler,KE,Paredes,KS,Cazalilla,Yip}.

In this Letter, we study a commensurate two-component bosonic
system described by the on-site Hubbard Hamiltonian:
\begin{equation}
H =  - \! \!  \sum_{< ij > \, \sigma}  (t_{\sigma}
a^{\dagger}_{i\sigma} a^{}_{j\sigma} + {\rm H.c.}) +
{1\over 2} \sum_{i \, \sigma \sigma'}  U_{\sigma \sigma'}   \,
n_{i\sigma} n_{i\sigma'}  \, . \label{Ham}
\end{equation}
Here $a^{\dagger}_{i\sigma}$ creates a boson of the sort $\sigma =
A,B$ on site $i$,
$n_{i\sigma}=a^{\dagger}_{i\sigma}a^{}_{i\sigma}$, and $<ij
>$ denotes pairs of nearest-neighbor sites. In what follows, we
consider only  equal filling factors of the components,
$n_A=n_B=n$, with $n$ integer, and, for brevity, denote
$U_{AB}=-V$, $U_{\sigma \sigma}=U_{\sigma}$.
Similar (incommensurate) two-species {\it bosonic} model
has been studied recently to look at the differences with
the {\it fermionic} Hubbard model \cite{Massimo}.

At double commensurate filling, recent mean-field analysis of the
model (\ref{Ham}) in \cite{Chen} failed to reveal phases and,
correspondingly, phase transitions which can not be reduced to 
simple mixtures of single-component states. This conclusion is
disappointing considering predictions of other strongly correlated
superfluid groundstates for two-component inconvertible bosons: a
paired superfluid vacuum (PSF), which is equivalent to the
superfluid state of diatomic molecules and to BCS superconductor
\cite{KE,Paredes,Cazalilla,KPS}; and a super-counter-fluid (SCF),
in which the net atomic superfluid current is zero, and yet the
equal currents of the components in opposite directions are
superfluid \cite{KS,KPS}.

In this Letter, we perform MC simulations of the
$(d+1)$-dimensional classical analog of the
on-site Hubbard model (\ref{Ham}) and find
five stable superfluid and insulating phases:
(i) MI, (ii) MI of sort A and SF of sort B ($\rm{MI_B + SF_A}$)
and its A $\leftrightarrow$ B analog, (iii) SF of sort A and SF of
sort B (2SF), (iv)  PSF, and (v) SCF. An interacting mixture of
two mutually penetrable superfluids (2SF) exists even without the
optical lattice and corresponds to the $t\gg U_{\sigma , \sigma
'}$ limit. This state has two non-zero complex order-parameters
$\langle \psi_A \rangle$ and $\langle
\psi_B\rangle$. By increasing either $U_B$ or $U_A$ one drives
the corresponding component from the superfluid
to the Mott-insulating state; accordingly, in $\rm{MI_B + SF_A}$
we have finite $\langle \psi_A \rangle$, and zero $\langle \psi_B
\rangle$. When both $U_A$ and $U_B$ are  strong, the groundstate
is MI with all order parameters being zero. Our proof
then concerns the existence of PSF and SCF phases; in both phases
$\langle \psi_A \rangle = \langle \psi_B \rangle = 0$, while
$\Phi_{PSF}=\langle \psi_A \psi_B \rangle \ne 0$ in PSF, and
$\Phi_{SCF}= \langle \psi_A \psi^{\dagger}_B \rangle \ne 0$ in
SCF. It is worth noting that, while the PSF, representing
atomic A+B pairing,  requires ~$V>0$,
the SCF describes pairing
of particles  A and holes B
and
occurs when ~$V<0$.

The most surprising MC result is that 2SF--MI transition may be
I-order. This result also follows from the
mean field (MF) analysis of the problem along the lines suggested
in \cite{Sachdev} for the single-component case. Finally,
we develop microscopic arguments explaining why the I-order
transition is generic for models with pairing interactions,
and show that MI groundstates may be further
classified in terms of their excitation spectrum.

To prove that possible groundstates of  Eq.~(\ref{Ham}) include
PSF, we assume the inter-exchange symmetry $A \longleftrightarrow
B$, implying $t_A = t_B = t$ and $U_{A}=U_{B}=U$, and consider the
limit described by two strong inequalities: $t/U \ll 1$ and
$\gamma/U \ll 1$, with $\gamma = U-V$ [here $V > 0$ and $\gamma >
0$; at $\gamma < 0$ the system collapses.] Then, the effective
low-energy Hilbert space is determined by states were on each site
$n_{iA}=n_{iB}$; these are separated from other states by a large
pair-breaking gap $\approx U$. We thus naturally arrive at the
description in terms of {\it pairs}. In the second-order
perturbation theory in $t/U$ (cf., e.g., \cite{KS})  the dynamics
of pairs is given by the effective Hamiltonian (we omit terms
proportional to the total number of particles):
\begin{eqnarray}
H_{\rm p} = -\tilde{t} \sum_{< ij >} [(O_i^{-}O_j^{+} + {\rm H.c.}) +
2 m_i m_j] + \gamma \sum_i m_i^2 \, .
\label{Ham_p}
\end{eqnarray}
Here $m_i$ are pair occupation numbers, the raising
operator  $O_i^{+}$ is defined by $\langle m_i' |O_i^{+} | m_i
\rangle = (m_i+1) \, \delta_{m_i',m_i+1}$,
$O_i^{-}=(O_i^{+})^{\dagger}$, and $\tilde{t}=2t^2/U$.
In contrast to the standard single-boson hopping
that scales linearly with the typical occupation number
the hopping amplitude for pairs is quadratic in $m_i$.
If potential energy terms in Eq.~(\ref{Ham_p}) were omitted,
the groundstate would collapse to a droplet with the
diameter comparable to the lattice constant.
The second term in the brackets, Eq.~(\ref{Ham_p}),
 describes nearest-neighbor
attraction, and further enhances collapse instability. A stable
groundstate arises only when the on-site repulsion
$\gamma $ is strong enough.
On the other hand, at very large $\gamma$ the commensurate groundstate is MI.
The question is then
whether for some $\tilde{t}/\gamma$, the groundstate is PSF rather
then MI or collapsed. The positive answer is readily seen in the
limit of very large molecular filling factor $m=n/2 \gg 1$. In the
region $\tilde{t} \ll \gamma \ll m^2 \tilde{t}$ the system is
stable against collapse; the nearest-neighbor attraction is
negligible. Since the maximum insulating gap $\sim \gamma$ does
not depend on $m$ \cite{Fisher}, we conclude that for $\gamma <
m^2 \tilde{t} $ the groundstate must be superfluid. [Using
mapping to the quantum rotor model \cite{Sachdev} we know, in fact,
that MI state requires $\gamma > m^2\tilde{t}$].
Finally, for $m=1$ we have performed quantum MC simulations of
Eq.~(\ref{Ham_p}) in $d=2$ and found that at $ \gamma = 10 \tilde{t}$
the groundstate is superfluid.
\begin{figure*}[thb]
\hspace*{0.15cm} \includegraphics[width=4.8cm]{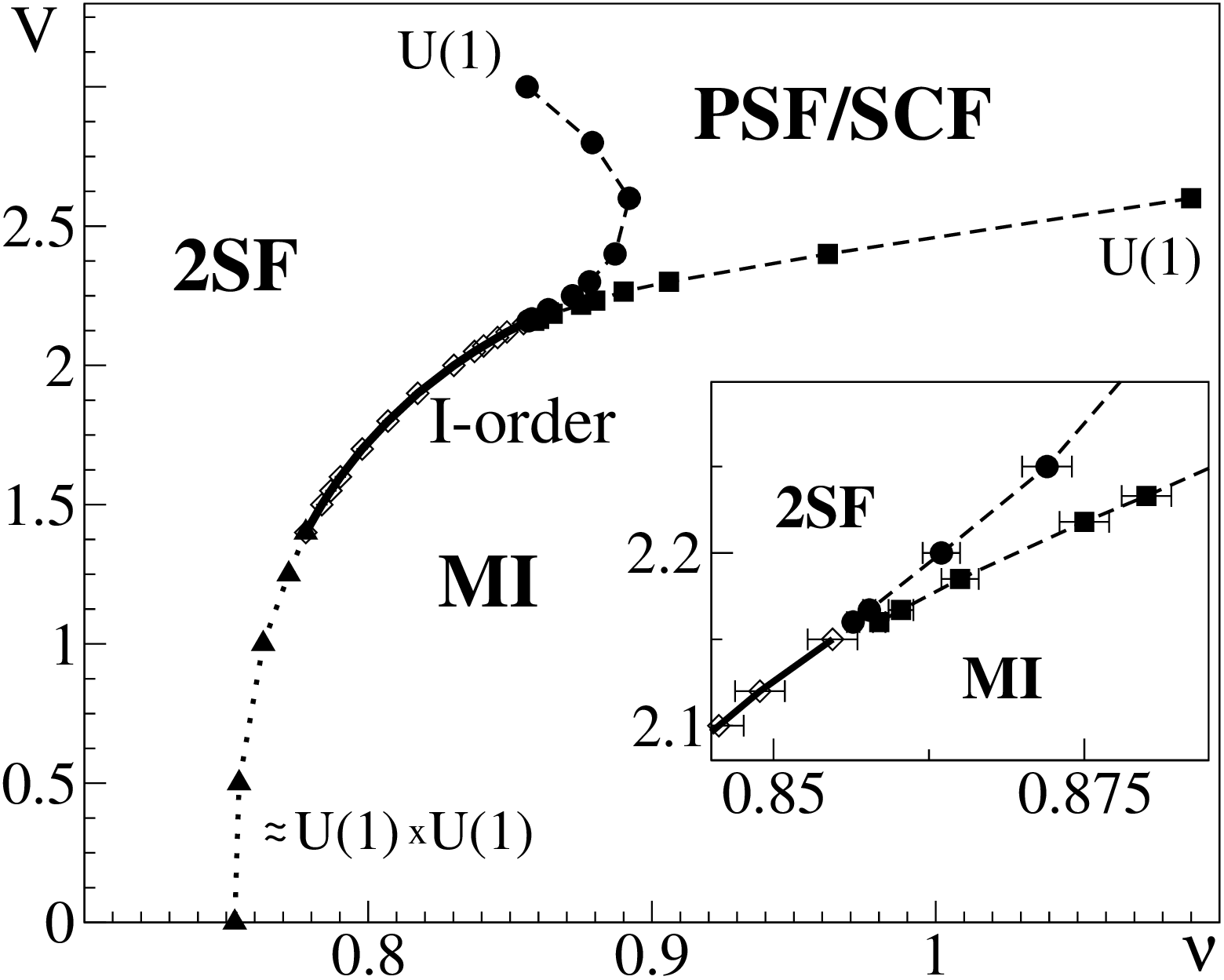}
\hspace*{1.cm}   \includegraphics[width=4.8cm]{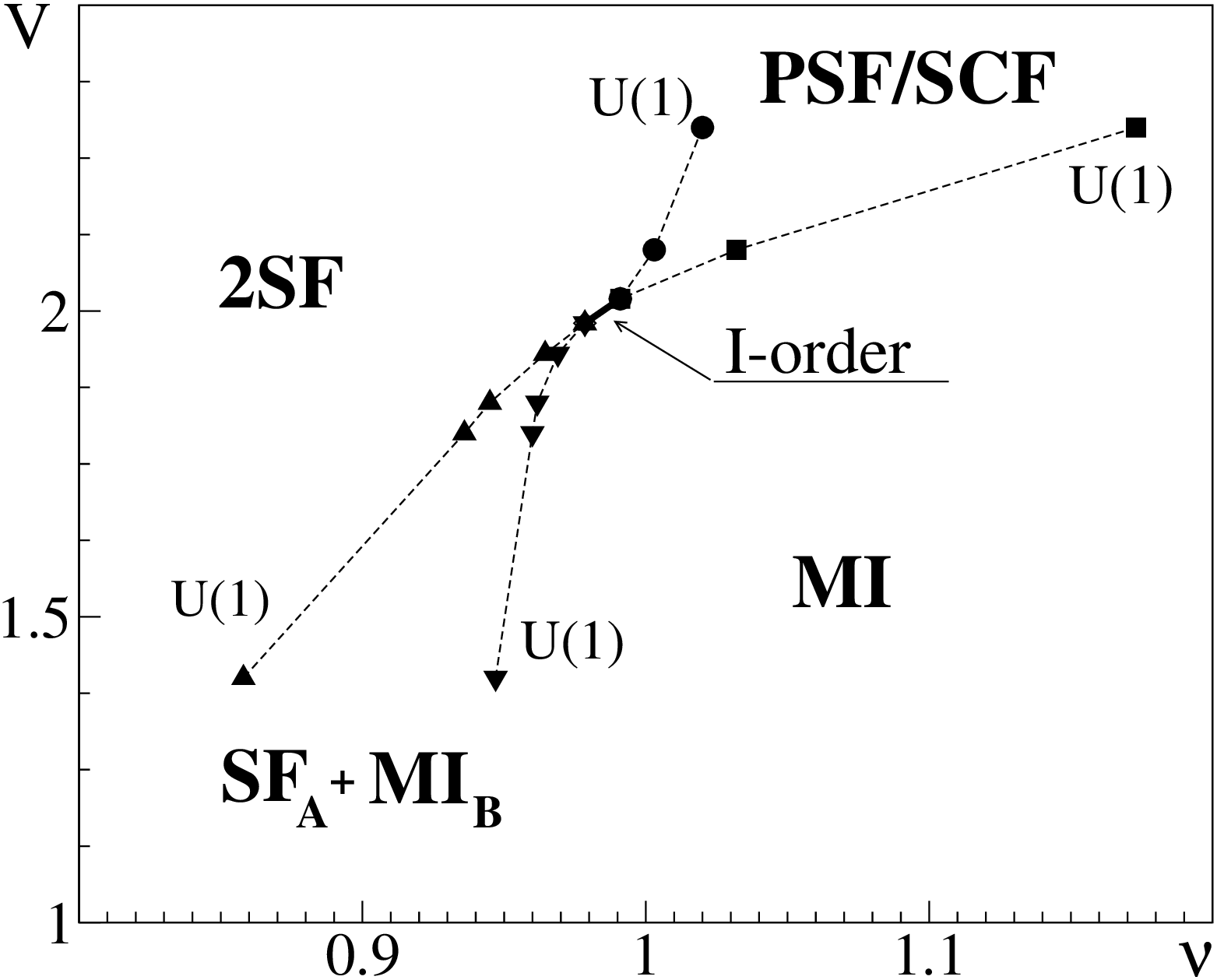}
\hspace*{1.cm}   \includegraphics[width=4.8cm]{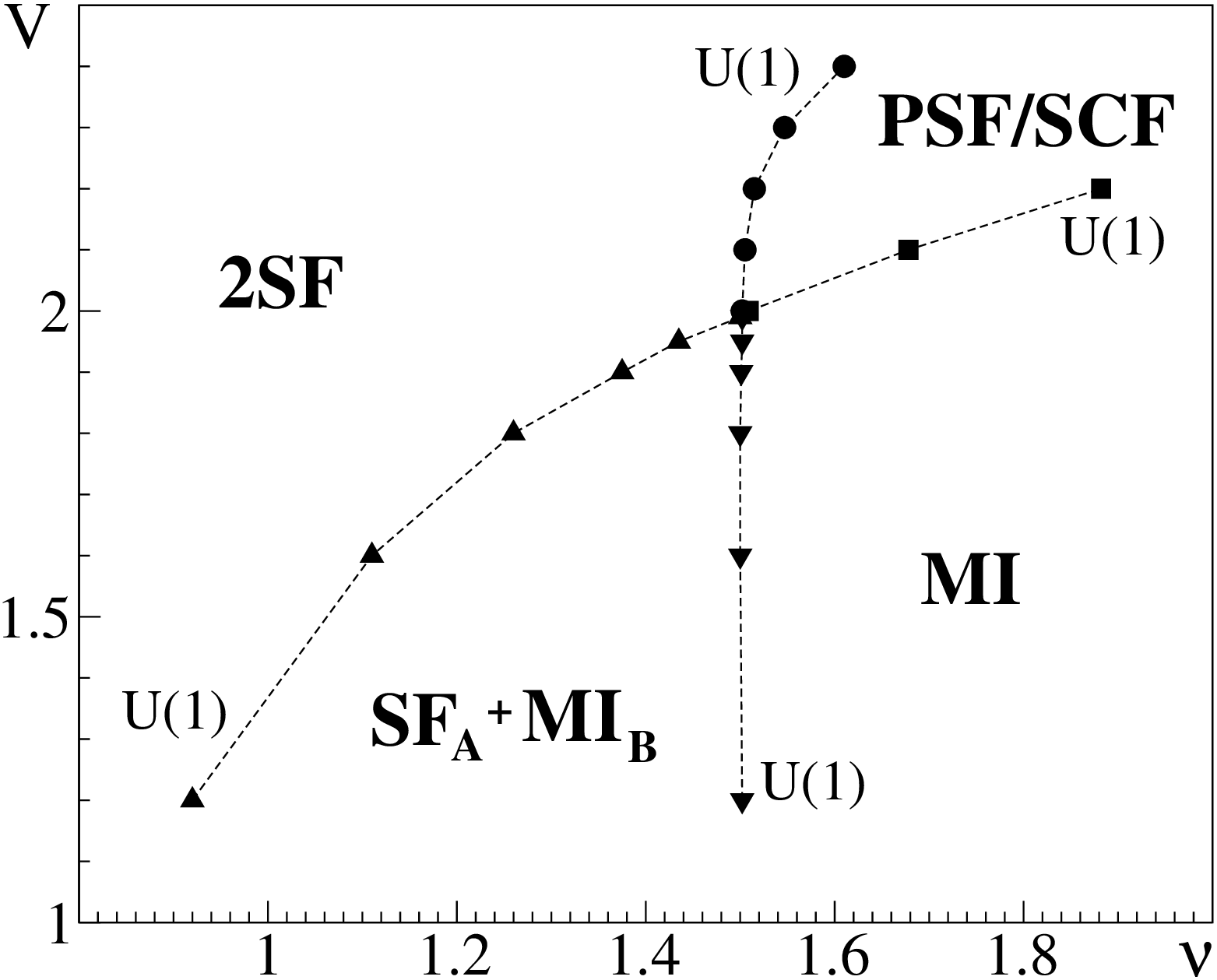}
\vspace*{-1.7cm} \caption{ Phase diagrams of the $d=2$ two-component
$J$-current model in the $(V,\nu )$-plain for the symmetric,
$U_A=U_B=2$ (left), slightly asymmetric, $U_A=1.6$, $U_B=2$
(center), and strongly asymmetric, $U_A=1$, $U_B=2$ (right)
models. The $1^{\rm st}$-order phase transition line is
dramatically reduced in the presence of weak anisotropy, and
completely disappears for strong anisotropy between the
components. All horizontal errorbars are smaller than point sizes
(typically of order $10^{-3}$), and lines are used to guide the
eye and to distinguish between different phase boundaries.
The insert shows more clearly the region where 2SF, PSF and MI phases meet.
(Commensurability and intrinsic symmetry of the $J$-current model
result in a straightforward mapping of the SCF regime onto PSF one: $V
\to -V$.)  } \label{Sdiagram}
\end{figure*}

Similarly, the existence of the SCF phase can be shown
in the limit $t/U \ll 1$ and $|\tilde{\gamma}|/U \ll 1$
[where $\tilde{\gamma} = U+V$ and $V$ is negative] studied in
Ref.~\cite{KS}. The effective Hamiltonian can be now written in terms
of the spin-$S=(n_A+n_B)/2$ operators, $H_{\rm
S} = -\tilde{t} \sum_{< ij >} {\bf S}_i {\bf S}_j + \tilde{\gamma}
\sum_i (S_{iz})^2$, which for small positive $\tilde{\gamma}$
has the easy-plane ferromagnetic groundstate, or SCF \cite{KS}.
Furthermore, one can show that PSF and SCF are qualitatively
similar and SCF may be
viewed as the result of pairing between particles of one component
and holes of another \cite{KPS}. {\it Because of this equivalence,
we discuss below only the $V>0$ case, that is, the
case of the PSF.}

To reveal the global structure of the phase diagram we performed
Monte Carlo simulations for the $(d+1)$-dimensional classical
analog of the bosonic Hubbard model. The so-called J-current model
\cite{Wallin94} is built on particle worldlines (space-time currents)
in discrete imaginary time, and we straightforwardly generalize
it to the two-component case:
\begin{equation}
S =  \sum_{\sigma ,\sigma '} \sum_i \: \tilde{U}_{\sigma \sigma '}
\: {\vec J}^{\,(\sigma)}_{i} \cdot {\vec J}^{\,(\sigma ')}_{i}  \,
. \label{Class}
\end{equation}
Here  ${\vec J}^{\,(\sigma )}_{i} $ are integer-value currents
[$(d+1)$-dimensional vectors] subject to the local zero-divergence
constraint, $\nabla \cdot {\vec J}^{\,(\sigma )}_i =0$, and
$\tilde{U}_{\sigma \sigma '} \sim U_{\sigma \sigma '}/t$ relate
the effective action parameters to the original Hubbard
Hamiltonian. This model has the same superfluid and insulating
phases as Eq.~(\ref{Ham}), and we use it to understand the
topology of phase boundaries, the existence of multicritical
points, and first-order lines. We find it convenient to fix $U_A$
and $U_B$ and to plot results in the $(V, \nu )$-plane, where $\nu
\sim  1/t$ is the scaling factor for all three dimensionless
parameters. The superfluid phases are identified by looking at
various superfluid stiffnesses, $\rho_s^{(\sigma )}=\langle [{\vec
W}^{(\sigma )}]^2\rangle /d L^{d-2} $, $\rho_s^{(PSF/SCF
)}=\langle [{\vec W}^{(A )} \pm {\vec W}^{( B)} ]^2\rangle
/D L^{D-2} $, expressed in terms of the winding number
fluctuations \cite{Ceperley}, where ${\vec W}^{(\sigma )}  =L^{-1}
\sum_i {\vec J}^{\,(\sigma )}_i$ (superfluid stiffness and
compressibility are equal in the space-time symmetric model).

In Fig.~\ref{Sdiagram}, we present the phase diagram of the
two-component J-current model in $(d+1=3)$-dimensions. 
Corresponding
superfluid stiffness goes to zero continuously when approaching
the lines of the critical points labeled as U(1). 
The correlation radius exponent (obtained
from finite-size corrections) is consistent with the known value
for the U(1) universality class in 3D. The first-order transition
was identified by (i) double-peak structure of the energy
distribution function (in small-size systems), and (ii) hysteresis
loops in all quantities (in large-size systems). Though we have
not performed exhaustive MC study of the phase diagram in other
dimensions we found (i) the I-order 2SF--MI transition in $d=3$,
and (ii) no evidence for the first-order 2SF--MI transition in
$d=1$.

The I-order 2SF--MI
line in the symmetric case ( $U_A=U_B$), becomes strongly
suppressed by the anisotropy, $U_A-U_B \ne 0$, between the
components. For $U_B/U_A = 2$, the point where all four phases meet
is already a simple cross of two U(1) lines --- decoupled $U(1)
\times U(1)$ tetracritical point (see, e.g., \cite{Aharony}).
Points where the I-order line starts and ends represent
multicritical points.

\begin{figure}[tbp]
\vspace*{0.2cm}
\includegraphics[width=6.5cm]{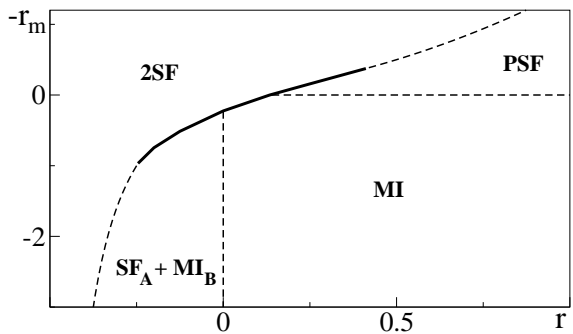}
\vspace*{-4.2cm} \caption{The mean-field diagram for $r_A=r$, $r_B=r+0.5$
The bold solid line corresponds to the first-order transition, and
dashed lines describe continuous transitions. In the limit of $r_B
\to r_A$, the $\rm{SF_A+MI_B}$ domain vanishes.} \label{MF}
\end{figure}

Normally, first-order transitions can be qualitatively accounted
for in simple mean-field models, and we propose such a model for
our case. Away from the multicritical region, all transitions are
of the $U(1)$-universality class and, thus, described by the
corresponding $|\psi|^4$ actions \cite{Fisher} for atomic,
$\psi_A$, $\psi_B$, and molecular, $\Phi$, fields. We arrive at
the simplest effective free energy by combining three $|\psi|^4$
actions and writing the interaction term in the form of the
molecule ``creation/annihilation" process out of A- and
B-particles. Omitting gradient terms:
\begin{eqnarray}
{\cal F} & = & {1 \over 2}
\big [ r_A|\psi_A|^2 + r_B|\psi_B|^2 + r_M|\Phi|^2 \big] + {1 \over 4}
\big [g_A |\psi_A|^4 \nonumber \\
& + &  g_B|\psi_B|^4 + g_M|\Phi|^4  \big]
- g (\Phi^* \psi_A \psi_B + \rm{c.c.}) \; . \label{Action}
\end{eqnarray}

The mean-field phase diagram follows from minimization of ${\cal
F}$. In Fig.\ref{MF}, taking advantage of the scaling freedom for all the fields and
${\cal F}$, we set $g_a=g_B=g_M=g=1$.
It reproduces correctly the topology of boundaries between
the phases and some of their properties. If the A-B asymmetry is
not large, it features a I-order line, see Fig.~\ref{MF}. In the
strongly anisotropic case, $|r_B-r_A|>1$, the MF theory also
captures the disappearance of the I-order transition. On another
hand, the prediction of the pronounced I-order 2SF--PSF line does
not agree with the numerical data in 3D. With our system sizes up
to $128^3$ sites we did not find any evidence for the I-order
2SF--PSF or 2SF--($\rm{MI_B + SF_A}$) transitions, see insert in
Fig.~\ref{Sdiagram}. It is probably too early to draw the final
conclusion on the structure of the multicritical point, because a
similar study in 4D (for the system size $32^4$) revealed a tiny
(but finite) I-order 2SF-PSF line. In any case, the suppression of
the  I-order 2SF-PSF transition constitutes a strong deviation
from the MF prediction.

{\it First-order SF--MI transition in the single-component
system.} It is generally accepted that in the single-component,
commensurate Bose system the SF--MI transition is continuous
\cite{Fisher}. Numerous simulations of the on-site Hubbard and
J-current models perfectly agree with this picture (for the
latest simulation see \cite{AS}).

Excitations in MI are gapped and described as quasiparticles and
quasiholes with the relativistic dispersion law at small momenta
(for small gaps): $\epsilon ({\bf k}) = \sqrt{\Delta^2 + c^2 k^2}$,
where $c$ is the velocity of sound in the SF phase.
The dilute gas of quasiparticles is characterized by
the effective mass $m_* = \Delta /c^2$ and some s-wave scattering
amplitude, $a_*$ (to be specific, we assume that $d=3$). If the
scattering length is positive, the state of the dilute excitation
gas with density $n_{\rm qp}$ is stable, and the energy density
cost of creating it is $E_{\rm qp}=(\Delta - \mu )n_{\rm qp}
+ (2\pi a_*/m_*)n_{\rm qp}^2$, where $\mu $ is the chemical potential.
Since the effective longwave action for the U(1)-transition
has {\it positive} coefficient in front of the $|\psi |^4$ term,
in the vicinity of the critical point the MI state is always described
by positive $a_*$. As the chemical potential is increased above the
threshold value $\Delta$, the system state becomes superfluid
(this continuous MI-SF transition, induced by adding extra particles/holes,
is mean-field like \cite{Fisher}).

Imagine now a MI state with gaped quasiparticle excitations, but
now with {\it negative} effective scattering length. Although the
MI vacuum itself may remain stable, the state of the quasiparticle
gas at any small density $n_{\rm qp}|a_*|^3 \ll 1 $ is unstable
against collapse to a dense droplet. We thus conclude that this MI
will undergo a first-order phase transition to the superfluid
state at some $\mu = \mu_c$ finite distance below $\Delta$ to gain
negative potential energy. Furthermore, if for some system
parameters $\Delta \ne 0$, but $\mu_c =0$, the MI--SF transition
in the commensurate system will happen by I-order scenario too.

\begin{figure}[tbp]
\vspace*{0.2cm}
\includegraphics[width=6.5cm]{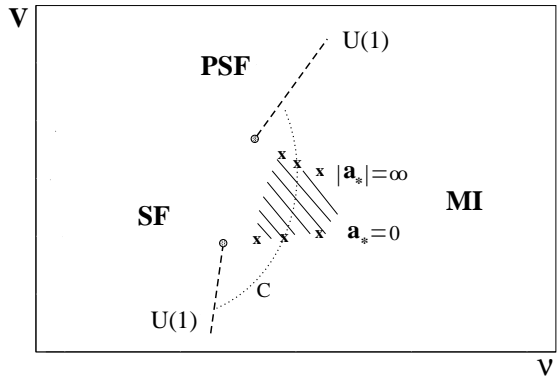}
\vspace*{-3.5cm} \caption{Sketch of possible MI phases in the
commensurate, $n=$even, single-component Hubbard model with
pairing interactions and phase transition lines from MI to the SF
and PSF phases. The region between crosses is characterized by
first-order MI--SF transition as a function of $\mu$.}
\label{MIphases}
\end{figure}

We now argue that MI states with negative $a_*$ naturally arise in
models with strong pairing interactions when potential energy
favors two bosons on the same site, but increases fast for
occupation numbers $n_i>2$ to prevent collapse. For sufficiently
strong pairing, one may have a superfluid state of tight molecules,
or PSF (cf. \cite{KKS}). When repulsion between molecules is
increased, PSF undergoes a standard continuous PSF--MI transition.
In the vicinity of the transition point the lowest excitations
above MI are {\it bound} bosonic pairs. We observe then, that
depending on the value of the pairing interaction there should
exist at least three MI groundstates distinguished by the value of
the effective scattering length of its quasiparticles: as we go
along the line $C$ in Fig.~\ref{MIphases}, $a_*$ starts from
positive value (the quasiparticle gas is stable), then changes
sign and becomes negative (the quasiparticle gas is unstable
against collapse), and finally goes through the pole and changes
sign again (the quasiparticle gas of molecules is stable). It
seems unlikely that SF, PSF, and three different MI phases meet at
the same point. In our view, the intersection of the SF--MI and
$|a_*|=0$ lines marks the beginning of the first-order SF--MI
transition, while the intersection of the PSF--MI and
$|a_*|=\infty $ lines marks its end. In this picture, the critical
point where the SF--MI line changes from continuous to first-order
is characterized by the continuous Lorentz-invariant action with
zero $|\psi |^4$ term.

In $d=2$, the weak logarithmic dependence of $a_*$ on
quasiparticle density does not change the qualitative picture,
because the first-order transition involves finite particle
density jumps. In $d=1$, the notion of the scattering length is
ill-defined, and two quasiparticles in the longwave limit either
form a bound state or repel each other like hard-core spheres. We
conjecture then, that in $d=1$ (i) MI with first-order transition
in $\mu$ does not exist, (ii) the SF--MI transition is always
continuous.

The above considerations readily generalize to the A-B symmetric
two-component case. Now, the criterion for the MI groundstate,
which is unstable against light doping by A- and B-particles (for
the symmetric case $\Delta_A =\Delta_B =\Delta $), follows from the
scattering matrix $ (a_*)_{\sigma  \sigma'} $ becoming
non-positive definite due to increasing attraction between the
components. We note, that this criterion {\it must} be satisfied
in a finite region in parameter space since existence of
AB-molecules implies that $ (a_*)_{AB} $ can be arbitrarily large
and negative before going through the pole corresponding to the
formation of the bound state. This consideration, in complete
analogy with the single-component case, suggests that PSF--MI and
2SF--MI lines are ``bridged'' by the first-order line in agreement
with MC simulations and MF analysis.

To explain the suppression and disappearance of the first-order
region when the symmetry between the A and B components is broken
(see Fig.~\ref{Sdiagram}) we suggest that, for strong anisotropy,
the lowest excitations above the MI groundstate in the whole
parameter range are either A-particles or  AB-molecules, and there
is no reason for collective instability in the quasiparticle gas.
Formally, this corresponds to pushing the $ |(a_*)_{AB}|=\infty $
line into the $\rm{SF_A+MI_B}$ phase --- this possibility does not
exist in the symmetric case.

We are grateful to A.~Patashinskii and S.~Sachdev for  fruitful
discussions and valuable comments. This work was supported by the
National Science Foundation and PSC CUNY grants. B.S. acknowledges
also a support from the Netherlands Organization for Scientific
Research (NWO).

\end{document}